\documentclass[amsmath,amssymb,twocolumn]{revtex4}

\usepackage{graphicx,hyperref}
\usepackage[varg]{txfonts}

\newcommand{\ii}{\mathrm{i}}
\newcommand{\e}{\mathrm{e}}
\renewcommand{\d}{\mathrm{d}}

\newcommand{\cvector}[1]{\left(\begin{array}{c}#1\end{array}\right)}
\renewcommand{\matrix}[2]{\left(\begin{array}{#1}#2\end{array}\right)}

\begin{document}

\title{Trajectories of point particles in cosmology and the Zel'dovich approximation}
\author{Matthias Bartelmann}
\affiliation{Heidelberg University, Zentrum f\"ur Astronomie, Institut f\"ur Theoretische Astrophysik, Philosophenweg 12, 69120 Heidelberg, Germany}

\begin{abstract}
Using a Green's function approach, we compare the trajectories of classical Hamiltonian point particles in an expanding space-time to the effectively inertial trajectories in the Zel'dovich approximation. It is shown that the effective gravitational potential accelerating the particles relative to the Zel'dovich trajectories vanishes exactly initially as a consequence of the continuity equation, and acts only during a short, early period. The Green's function approach suggests an iterative scheme for improving the Zel'dovich trajectories, which can be analytically solved. We construct these trajectories explicitly and show how they interpolate between the Zel'dovich and the exact trajectories. The effective gravitational potential acting on the improved trajectories is substantially smaller at late times than the potential acting on the exact trajectories. The results may be useful for Lagrangian perturbation theory and for numerical simulations.
\end{abstract}

\maketitle

\section{Introduction}

An analysis of the trajectories of classical, gravitating point particles in an expanding (Friedman-Lema\^\i tre-Robertson-Walker) space-time quickly leads to the conclusion that the comoving displacement of particles from their initial positions is finite in reality even in the limit of infinite times. In sharp contrast, the remarkably successful Zel'dovich approximation \cite{1970A&A.....5...84Z} asserts that particle trajectories are well approximated by trajectories which resemble inertial motion in a suitable time coordinate, which manifestly leads to unbounded displacements. How can these two approaches be reconciled?

Beginning with the retarded Green's functions of classical point particles in a static and in an expanding space-time, and suitably re-grouping the terms in the point-particle Hamiltonian in an expanding space-time, we derive the effective gravitational potential acting on point particles relative to free Zel'dovich trajectories. The result shows that, while the density field is evolving linearly, this effective potential acts only for a short period of time at early cosmic times. This is a direct consequence of the relation between the initial density contrast and the initial particle velocities enforced by the matter continuity equation. This result may contribute to clarifying the astounding success of the Zel'dovich approximation.

The Green's function approach further allows the construction of an iterative scheme for deriving free particle trajectories in an expanding space-time, which can be solved analytically. These newly derived, improved trajectories interpolate between the free trajectories of the exact cosmological Hamiltonian and the Zel'dovich trajectories. They offer several advantages compared to the Zel'dovich approximation as well as compared to the trajectories of the exact Hamiltonian: 

Compared to Zel'dovich, the new trajectories lead to a substantial reduction of the unwanted re-expansion of structures after shell crossing. Compared to the trajectories of the exact Hamiltonian, the effective gravitational potential acting on the newly derived trajectories decays much faster in time than the Newtonian gravitational potential does. For numerical simulations, this may allow to achieve a given spatial and temporal resolution with substantially fewer time steps. The new trajectories may also be helpful for extending Lagrangian perturbation theory (see e.g.~\cite{1995A&A...296..575B, 1992MNRAS.254..729B, 1994MNRAS.267..811B, 1993MNRAS.264..375B, 1997GReGr..29..733E, 2008PhRvD..78h3503B} and \cite{2002PhR...367....1B} for a review).

We describe the Green's function approach in Sect.~2 and apply it to an analysis of the Zel'dovich approximation in Sect.~3. In Sect.~4, we construct the iterative scheme leading to the new free trajectories, and we summarise our results in Sect.~5.

\section{Particle trajectories}

\subsection{Particles in a static space-time}

With the Hamilton function $\mathcal{H} = \vec p^{\,2}/(2m)$ of a classical free point particle with mass $m$ in a static space-time, the equations of motion for a free particle read
\begin{equation}
  \partial_tx = \mathcal{J}\partial_x\mathcal{H} = \mathcal{K}x \;,\quad
  \mathcal{K} = \matrix{cc}{0 & m^{-1}\mathcal{I}_3 \\ 0 & 0 \\}\;,
\label{eq:02-1}
\end{equation}
where $x = (\vec q, \vec p\,)$ is the particle position in phase space. The $6\times6$ dimensional matrix $\mathcal{K}$ is called \emph{force matrix}, $\mathcal{J}$ is the usual symplectic matrix
\begin{equation}
  \mathcal{J} = \matrix{cc}{0 & \mathcal{I}_3\\ -\mathcal{I}_3 & 0}
\label{eq:02-2}
\end{equation}
and $\mathcal{I}_n$ is the $n$-dimensional unit matrix.

The retarded Green's function of the Hamiltonian equations (\ref{eq:02-1}) is
\begin{equation}
  \bar G_\mathrm{R}(t,t') =
  \matrix{cc}{\mathcal{I}_3 & m^{-1}(t-t')\mathcal{I}_3 \\ 0 & \mathcal{I}_3 \\}\theta(t-t')\;;
\label{eq:02-3}
\end{equation}
see Appendix~\ref{app:A}. Beginning at the initial position $x^\mathrm{(i)} = (\vec q^\mathrm{\,(i)},\vec p^\mathrm{\,(i)})$ in phase space, the free solution $x^0(t)$ evolves as
\begin{equation}
  x^0(t) = \bar G_\mathrm{R}(t,t_\mathrm{i})x^\mathrm{(i)}
\label{eq:02-4}
\end{equation}
for $t\ge t_\mathrm{i}$. For a particle moving in a potential $v$, the potential gradient $\nabla_q v$ adds the inhomogeneity
\begin{equation}
  y(t) = \cvector{0\\-\nabla_qv}
\label{eq:02-5}
\end{equation} 
to the right-hand side of Hamilton's equations. Including this inhomogeneity, the phase-space trajectory is
\begin{equation}
  x(t) = \bar G_\mathrm{R}(t,t_\mathrm{i})x^\mathrm{(i)} -
  \int_{t_\mathrm{i}}^t\bar G_\mathrm{R}(t,t')\cvector{0\\\nabla_qv}\d t'\;.
\label{eq:02-6}
\end{equation} 

\subsection{Particles in an expanding space-time}

The effective Lagrange function for classical point particles in an expanding universe is
\begin{equation}
  L(\vec q, \dot{\vec q}, t) = \frac{m}{2}a^2\dot{\vec q}^{\,2}-m\phi\;,
\label{eq:02-7}
\end{equation}
where $\vec q$ are now \emph{comoving} spatial coordinates. The peculiar gravitational potential $\phi$ satisfies the Poisson equation
\begin{equation}
  \nabla_q^2\phi = 4\pi Ga^2\left(\rho-\bar\rho\right)\;;
\label{eq:02-8}
\end{equation}
see \cite{1980lssu.book.....P} and Appendix~\ref{app:B}.

We proceed by transforming the time coordinate $t$ to a dimension-less time coordinate $\tau = D_+-D_\mathrm{+}^\mathrm{(i)}$, where $D_+$ is the usual growth factor, i.e.\ the growing solution of the linear density perturbation equation. For convenience and without loss of generality, we define the time $\tau$ such that $\tau = \tau_\mathrm{i} = 0$ initially, when $D_+ = D_+^\mathrm{(i)} = 1$. We further normalise the cosmological scale factor $a$ such that $a_\mathrm{i} = a(\tau_\mathrm{i}) = 1$.

Since
\begin{equation}
  \d\tau = \d D_+ = \frac{\d a}{\d t}\frac{\d D_+}{\d a}\d t = HD_+f\d t\;,
\label{eq:02-9}
\end{equation}
time derivatives are related by
\begin{equation}
  \frac{\d}{\d t} = HD_+f\frac{\d}{\d\tau}\;,
\label{eq:02-10}
\end{equation}
with the usual definitions
\begin{equation}
  f := \frac{\d\ln D_+}{\d\ln a}\;,\quad H := \frac{\dot a}{a}\;.
\label{eq:02-11}
\end{equation}
First-order time derivatives thus transform as
\begin{equation}
  \dot{\vec q} = \frac{\d\vec q}{\d t} = HD_+f\frac{\d\vec q}{\d\tau} = HD_+f\vec q^{\,\prime}\;,
\label{eq:02-12}
\end{equation}
where the prime on the right-hand side denotes for now the derivative with respect to the new time coordinate $\tau$ rather than the cosmological time $t$.

This time transformation needs to leave the action unchanged, hence
\begin{align}
  S &= \int_1^2\d t\,L(\vec q, \dot{\vec q}, t) =
  \int_1^2\d\tau\,L'(\vec q, \vec q^{\,\prime}, \tau) \nonumber\\ &=
  \int_1^2\d t\,\frac{\d\tau}{\d t}\,L'(\vec q, \vec q^{\,\prime}, \tau)\;.
\label{eq:02-13}
\end{align}
With (\ref{eq:02-7}), (\ref{eq:02-10}) and (\ref{eq:02-12}), this requirement returns the effective Lagrange function
\begin{equation}
  L'(\vec q, \vec q^{\,\prime}, \tau) = \frac{\d t}{\d\tau}\,L(\vec q, \dot{\vec q}, t) =
  \frac{m}{2}a^2HD_+f\vec q^{\,\prime2} - \frac{m\phi}{HD_+f}\;,
\label{eq:02-14}
\end{equation}
where the new time coordinate is now $\tau$. Finally, we factorise the constant $mH_\mathrm{i}$ out of the effective Lagrange function, drop the prime on $L$ and replace $\vec q^{\,\prime}$ by $\dot{\vec q}$. Thus, from now on, we shall use
\begin{equation}
  L(\vec q,\dot{\vec q},\tau) = \frac{g(\tau)}{2}\dot{\vec q}^{\,2} - v(\vec q,\tau)
\label{eq:02-15}
\end{equation}
as the effective Lagrange function, with
\begin{equation}
  g(\tau) := a^2D_+fHH_\mathrm{i}^{-1}\;.
\label{eq:02-16}
\end{equation}
In the early universe, when the Einstein-de Sitter limit holds, we must have
\begin{equation}
  \left.f\right\vert_{\tau=0} = 1 \quad\mbox{and}\quad
  \left.g(\tau)\right\vert_{\tau=0} = 1\;.
\label{eq:02-17}
\end{equation}

The effective gravitational potential $v$ appearing in (\ref{eq:02-15}) is
\begin{equation}
  v(\vec q,\tau) := \frac{\phi}{HD_+fH_\mathrm{i}} = \frac{a^2\phi}{g(\tau)H_\mathrm{i}^2}
\label{eq:02-18}
\end{equation}
and thus obeys the Poisson equation
\begin{equation}
  \nabla_q^2v(\vec q,\tau) = \frac{4\pi Ga}{H_\mathrm{i}^2g(\tau)}\left(\rho-\bar\rho\right)
\label{eq:02-19}
\end{equation}
following from (\ref{eq:02-8}). Since the mean comoving cosmic matter density is
\begin{equation}
  \bar\rho = \frac{3H_\mathrm{i}^2}{8\pi G}\Omega_\mathrm{mi}
\label{eq:02-20}
\end{equation}
with the matter-density parameter $\Omega_\mathrm{mi}$ at the initial time, the Poisson equation (\ref{eq:02-19}) is
\begin{equation}
  \nabla_q^2v(\vec q,\tau) = \frac{3}{2}\frac{a}{g(\tau)}\Omega_\mathrm{mi}\delta\;.
\label{eq:02-21}
\end{equation} 

The canonically-conjugate momentum is
\begin{equation}
  \vec p = g(\tau)\dot{\vec q}\;,
\label{eq:02-22}
\end{equation}
leading to the Hamiltonian
\begin{equation}
  \mathcal{H} = \vec p\cdot\dot{\vec q}-L = \frac{\vec p^{\,2}}{2g(\tau)}+v(\vec q,\tau)
\label{eq:02-23}
\end{equation}
and the Hamiltonian equations of motion
\begin{equation}
  \dot{\vec q} = g^{-1}(\tau)\vec p \;,\quad
  \dot{\vec p} = -\nabla_qv\;.
\label{eq:02-24}
\end{equation}

Assuming a vortex-free initial velocity field, we can introduce a velocity potential $\psi$ such that
\begin{equation}
  \dot{\vec q}^\mathrm{\,(i)} = \nabla_q^\mathrm{(i)}\psi\;.
\label{eq:02-25}
\end{equation}
Note that $\psi$ must have the dimension of a length since $\tau$ is dimension-less. The continuity equation for the cosmic density evaluated at $\tau = 0$ with the initial velocity (\ref{eq:02-25}) requires
\begin{equation}
  \left.\dot\delta\right\vert_{\tau=0} =
  -\nabla_q^\mathrm{(i)}\cdot\dot{\vec q}^\mathrm{\,(i)} =
  -\left(\nabla_q^\mathrm{(i)}\right)^2\psi\;.
\label{eq:02-26}
\end{equation}
With $\delta = \delta_\mathrm{i}D_+ = \delta_\mathrm{i}(1+\tau)$ initially, we have $\left.\dot\delta\right\vert_{\tau=0} = \delta_\mathrm{i}$ and hence the Poisson equation
\begin{equation}
  \left(\nabla^\mathrm{(i)}\right)^2\psi = -\delta_\mathrm{i}
\label{eq:02-27}
\end{equation}
relating the initial density contrast to the velocity potential $\psi$. Since $g(\tau) = 1$ at $\tau = 0$, the initial conjugate momentum (\ref{eq:02-22}) is identical to the initial velocity,
\begin{equation}
  \vec p^\mathrm{\,(i)} = \dot{\vec q}^\mathrm{\,(i)} = \nabla_q^\mathrm{(i)}\psi\;.
\label{eq:02-28}
\end{equation}

The retarded Green's function solving the Hamiltonian equations (\ref{eq:02-24}) is
\begin{equation}
  G_\mathrm{R}(\tau, \tau') = \matrix{cc}{\mathcal{I}_3 & g_{qp}(\tau, \tau')\mathcal{I}_3 \\ 0 & \mathcal{I}_3 \\}\theta(\tau-\tau')
\label{eq:02-29}
\end{equation}
with
\begin{equation}
  g_{qp}(\tau, \tau') := \int_{\tau'}^\tau\frac{\d\bar\tau}{g(\bar\tau)}\;;
\label{eq:02-30}
\end{equation}
see Appendix~\ref{app:A}.

It is important to note that the propagator $g_{qp}(\tau, \tau')$ remains finite for $\tau\to\infty$ under realistic circumstances. In order to see this, we write
\begin{equation}
  g_{qp}(\tau,\tau') = H_\mathrm{i}\int_{\tau'}^\tau\frac{\d\bar\tau}{\bar a^2D_+Hf} =
  H_\mathrm{i}\int_{a'}^a\frac{\d\bar a}{\bar a^3H}\;,
\label{eq:02-31}
\end{equation}
where (\ref{eq:02-9}) was used to substitute the scale factor $a$ for $\tau$ as the integration variable. The ansatz $H = H_\mathrm{i}a^{-n}$ shows that the right-hand side of (\ref{eq:02-31}) is finite for $n<2$, which is satisfied for the matter-dominated era. For an Einstein-de Sitter universe,
\begin{equation}
  \lim_{\tau\to\infty}g_{qp}(\tau,\tau') = \frac{2}{\sqrt{1+\tau'}}\;.
\label{eq:02-32}
\end{equation}
The free spatial trajectory of a particle in an expanding space-time,
\begin{equation}
  \vec q^{\,0}(\tau) = \vec q^\mathrm{\,(i)} + g_{qp}(\tau,0)\vec p^\mathrm{\,(i)}\;,
\label{eq:02-33}
\end{equation}
shows that the particle can only travel by the finite amount
\begin{equation}
  \left\vert\vec q^{\,0}(\tau)-\vec q^\mathrm{\,(i)}\right\vert \le
  g_{qp}(\infty,0)\vec p^\mathrm{\,(i)}
\label{eq:02-34}
\end{equation}
even in an infinite time.

\section{Comparison to the Zel'dovich approximation}

In apparently sharp contrast to the result (\ref{eq:02-33}), the Zel'dovich approximation \cite{1970A&A.....5...84Z} asserts that the comoving particle trajectory $\vec q(\tau)$ is approximated by
\begin{equation}
  \vec q(\tau) = \vec q^\mathrm{\,(i)}+\tau\vec p^\mathrm{\,(i)}\;,
\label{eq:02-35}
\end{equation}
where $\vec p^\mathrm{\,(i)}$ is the initial conjugate particle momentum. This approximate inertial motion seems to be in conflict with the Hamiltonian (\ref{eq:02-23}) and with our previous conclusion that the Green's function for free Hamiltonian particles in an expanding space-time remains finite for $\tau\to\infty$.

In order to see how the Hamiltonian (\ref{eq:02-23}) can be reconciled with the Zel'dovich approximation, we re-write it in the form
\begin{equation}
  \mathcal{H} = \frac{\vec p^{\,2}}{2}+h(\tau)\frac{\vec p^{\,2}}{2}+v
\label{eq:02-36}
\end{equation}
with
\begin{equation}
  h(\tau) = g^{-1}(\tau)-1\;.
\label{eq:02-37}
\end{equation}
Since, as we saw before, $g(\tau)\to1$ for $\tau\to0$, the function $h(\tau)\to0$ initially, and the Hamiltonian then resembles that of a particle in static space-time.

We now treat the term $h\vec p^{\,2}/2$ in the Hamiltonian (\ref{eq:02-36}) as an \emph{inhomogeneity} in the equations of motion. According to the Hamiltonian equations, the inhomogeneity (\ref{eq:02-5}) then changes to
\begin{equation}
  y(\tau) = \cvector{h\vec p\\-\nabla_qv}\;.
\label{eq:02-38}
\end{equation}
Since the free Hamiltonian then equals that of a free particle in static space-time, we can write the solution in terms of the Green's function (\ref{eq:02-3}), with $t$ replaced by $\tau$ and the particle mass $m$ dropped,
\begin{equation}
  x(\tau) = \bar G_\mathrm{R}(\tau,0)x^\mathrm{(i)}+
  \int_0^\tau\bar G_\mathrm{R}(\tau,\tau')\cvector{h\vec p\\-\nabla_qv}\d\tau'\;.
\label{eq:02-39}
\end{equation}

In particular, the spatial trajectory $\vec q(\tau)$ is
\begin{equation}
  \vec q(\tau) = \vec q^\mathrm{\,(i)}+\tau\vec p^\mathrm{\,(i)}+\delta\vec q\;,
\label{eq:02-40}
\end{equation}
where
\begin{equation}
  \delta\vec q := \int_0^\tau\left(h\vec p-(\tau-\tau')\nabla_qv\right)\d\tau'
\label{eq:02-41}
\end{equation} 
quantifies the deviation from the Zel'dovich trajectory (\ref{eq:02-35}). By a partial integration in the first term on the right-hand side, we can re-write (\ref{eq:02-41}) as
\begin{align}
  \delta\vec q &= -\left.(\tau-\tau')h\vec p\,\right\vert_0^\tau \nonumber\\ &+
  \int_0^\tau(\tau-\tau')\left(\dot hp+h\dot p-\nabla_qv\right)\d\tau'\;.
\label{eq:02-42}
\end{align}
Since the boundary term vanishes and $\dot p = -\nabla_qv$, (\ref{eq:02-42}) shrinks to
\begin{equation}
  \delta\vec q = \int_0^\tau(\tau-\tau')
  \left(\dot h\vec p-g^{-1}(\tau')\nabla_qv\right)\d\tau'\;.
\label{eq:02-43}
\end{equation} 
Compared to the inertial Zel'dovich motion (\ref{eq:02-35}), the particle thus behaves as if it moved under the influence of an effective force
\begin{equation}
  \vec f = \dot h\vec p-g^{-1}(\tau')\nabla_qv\;.
\label{eq:02-44}
\end{equation}

Early in time, the momentum will be
\begin{equation}
  \vec p \approx \vec p^\mathrm{\,(i)} = \nabla_q^\mathrm{(i)}\psi
\label{eq:02-45}
\end{equation}
according to (\ref{eq:02-28}), where $\psi$ is the velocity potential introduced in (\ref{eq:02-25}). Using the Poisson equations (\ref{eq:02-21}) and (\ref{eq:02-27}), we can write
\begin{equation}
  \nabla_q^\mathrm{(i)}\psi = \int\frac{\d^3k}{(2\pi)^3}
  \frac{\ii\vec k}{k^2}\,\hat\delta_\mathrm{i}\,
  \e^{\ii\vec k\cdot\vec q^\mathrm{\,(i)}}
\label{eq:02-46}
\end{equation}
and
\begin{equation}
  \nabla_qv = -\frac{3}{2}\frac{aD_+}{g}\int\frac{\d^3k}{(2\pi)^3}
  \frac{\ii\vec k}{k^2}\,\hat\delta_\mathrm{i}\,
  \e^{\ii\vec k\cdot\left(\vec q^\mathrm{\,(i)}+\tau\vec p^\mathrm{\,(i)}\right)}
\label{eq:02-47}
\end{equation}
as long as the density contrast grows linearly, $\delta = D_+\delta_\mathrm{i}$.

The effective force is now
\begin{equation}
  \vec f = \int\frac{\d^3k}{(2\pi)^3}\frac{\ii\vec k}{k^2}\,\hat\delta_\mathrm{i}\,
  \e^{\ii\vec k\cdot\vec q^\mathrm{\,(i)}}
  \left\{\dot h+\frac{3}{2}\frac{aD_+}{g^2}\e^{\ii\vec k\cdot\tau\vec p^\mathrm{\,(i)}}\right\}\;.
\label{eq:02-48}
\end{equation}
At early times, the phase factor in (\ref{eq:02-48}) is near unity and the integrand approximated by
\begin{equation}
  \dot h+\frac{3}{2}\frac{aD_+}{g^2} = g^{-2}\left(\frac{3}{2}aD_+-\dot g\right)\;.
\label{eq:02-49}
\end{equation}

For an Einstein-de Sitter universe, $g = a^{3/2}$, further $D_+ = a = 1+\tau$, and
\begin{equation}
  g^{-2}\left(\frac{3}{2}aD_+-\dot g\right) = \frac{3}{2a}\left(1-a^{-3/2}\right)\;.
\label{eq:02-50}
\end{equation} 
This function drops to zero for $a\to 1$ and $a\to\infty$. It reaches a sharp maximum at
\begin{equation}
  a_\mathrm{max} = \left(\frac{5}{2}\right)^{2/3} \approx 1.84\;,
\label{eq:02-51}
\end{equation}
where it rises to $9/10(2/5)^{2/3} \approx 0.49$.

\begin{figure}
  \includegraphics[width=\hsize]{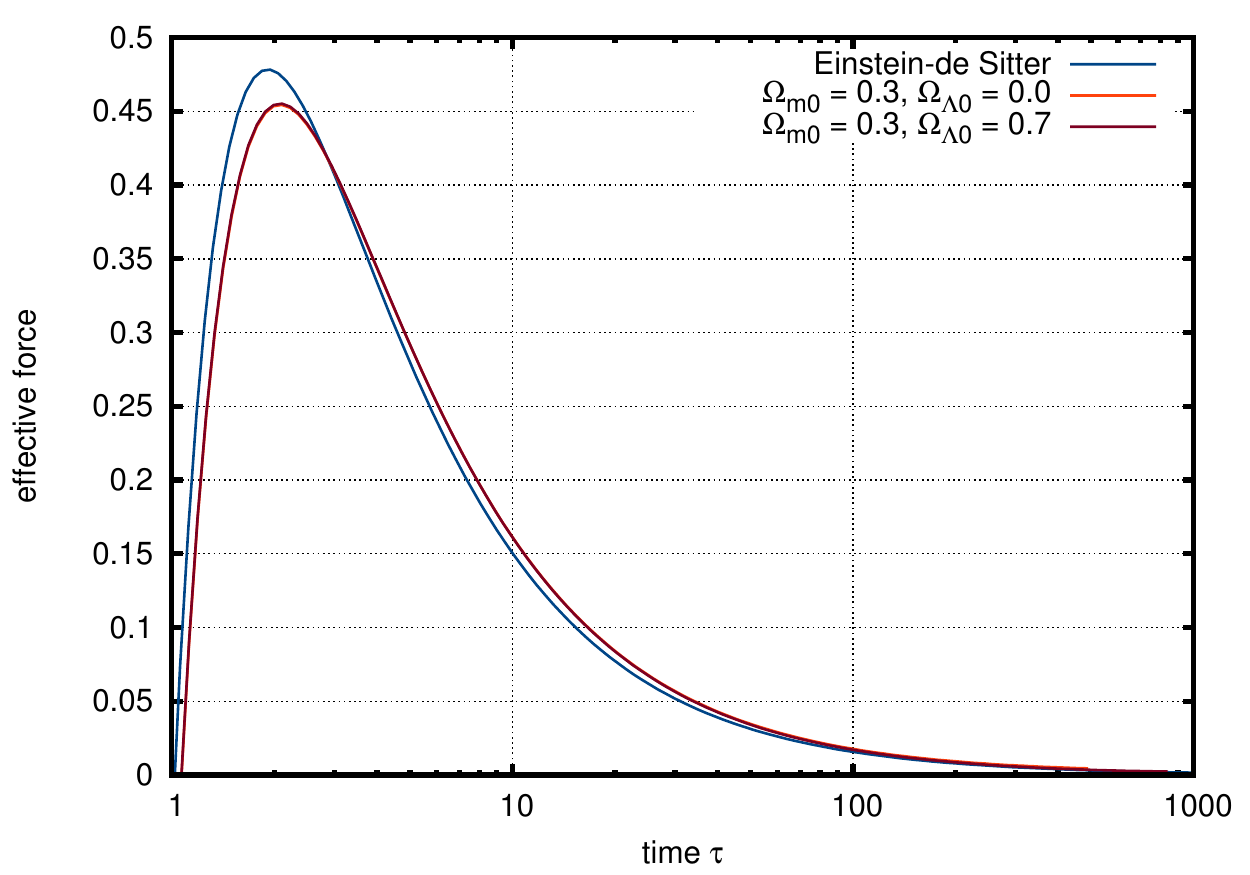}
\caption{Time evolution of the effective force $\vec f$ experienced by a point particle relative to the Zel'dovich trajectory.}
\label{fig:1}
\end{figure}

Even though these results were derived for the Einstein-de Sitter model, other cosmological models show a very similar behaviour because the Einstein-de Sitter limit is generally valid at early times (see Fig.~\ref{fig:1}).

We thus conclude that the effective force accelerating a point particle relative to the inertial Zel'dovich trajectories is small and acts only during a short time interval at early cosmic times. This justifies our evaluating the force field at early times only: at later times, it quickly drops to zero.

In addition, the expression (\ref{eq:02-48}) explicitly (but approximately) specifies the effective gravitational potential $v_\mathrm{eff}^\mathrm{(Z)}$ acting on a Zel'dovich trajectory. Its Fourier transform with respect to the Zel'dovich coordinates $\vec q = \vec q^\mathrm{\,(i)}+\tau\vec p^\mathrm{\,(i)}$ is
\begin{equation}
  \hat v_\mathrm{eff}^\mathrm{(Z)} = -\frac{\hat\delta_\mathrm{i}}{k^2}
  \left(
    \dot h\e^{-\ii\vec k\cdot\tau\vec p^\mathrm{\,(i)}}+\frac{3}{2}\frac{aD_+}{g^2}
  \right)\;.
\label{eq:02-52}
\end{equation} 

\section{Improved particle trajectories}

Equation (\ref{eq:02-43}) for the deviation from the Zel'dovich trajectory suggests to replace the inhomogeneity (\ref{eq:02-38}) by
\begin{equation}
  y(\tau) = \cvector{0\\\dot h\vec p-g^{-1}\nabla_qv}\;.
\label{eq:02-53}
\end{equation}
For free particles, $v = 0$. Then, according to (\ref{eq:02-6}), the remaining inhomogeneity implies the free solutions
\begin{equation}
  \vec q(\tau) = \vec q^\mathrm{\,(i)} + \tau\vec p^\mathrm{\,(i)} +
  \int_0^\tau(\tau-\tau')\dot h\vec p\,\d\tau'
\label{eq:02-54}
\end{equation}
for the spatial trajectory and
\begin{equation}
  \vec p(\tau) = \vec p^\mathrm{\,(i)} +
  \int_0^\tau\dot h\vec p\,\d\tau'
\label{eq:02-55}
\end{equation}
for the particle momentum. This is to say, free particles move in an expanding space-time as if they were freely moving in a static space-time with the momentum satisfying (\ref{eq:02-55}).

\begin{figure}
  \includegraphics[width=\hsize]{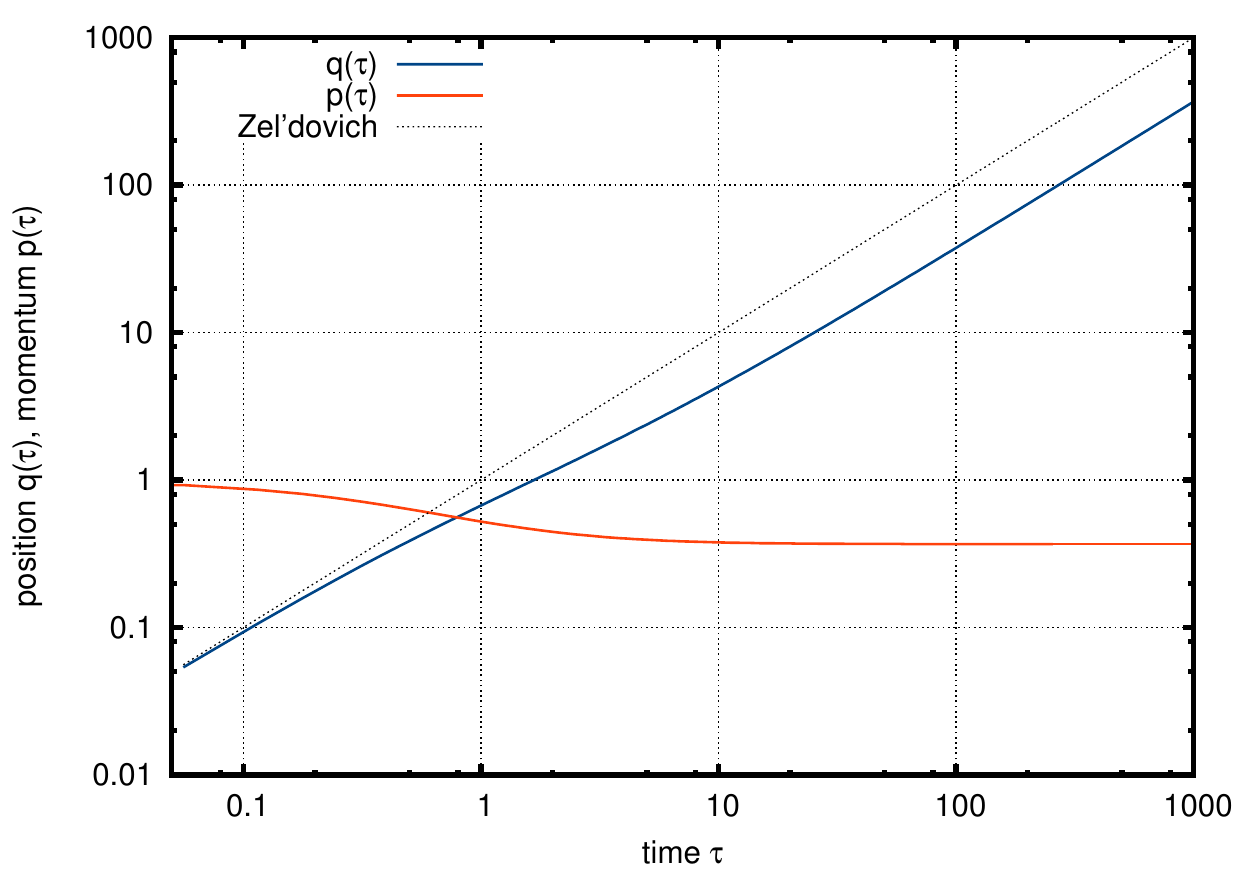}
\caption{The solutions (\ref{eq:02-55e}) for the free particle trajectories and (\ref{eq:02-55a}) for their conjugate momentum are shown here.}
\label{fig:2}
\end{figure}

The integral equation (\ref{eq:02-55}) for the momentum could be solved iteratively, beginning with inserting $\vec p = \vec p^\mathrm{\,(i)}$ into the integral as the zeroth-order solution. It can, however, easily be solved analytically after rewriting it as
\begin{equation}
  \vec p(\tau) = \vec p^\mathrm{\,(i)} + \int_0^{h(\tau)}\vec p\,\d h\;.
\label{eq:02-55a}
\end{equation}
Beginning with $\vec p = \vec p^\mathrm{\,(i)}$, it is easily seen that (\ref{eq:02-55a}) is solved by
\begin{equation}
  \vec p(\tau) = \vec p^\mathrm{\,(i)}\exp\left(h(\tau)\right)
\label{eq:02-55b}
\end{equation}
independent of the cosmological model. Inserting this solution into (\ref{eq:02-54}), we obtain the expression
\begin{equation}
  \vec q(\tau) = \vec q^\mathrm{\,(i)}+\vec p^\mathrm{\,(i)}\left(
    \tau+\int_0^\tau(\tau-\tau')\dot h\e^h\d\tau'
  \right)
\label{eq:02-55c}
\end{equation}
for the spatial trajectories. Writing
\begin{equation}
  \dot h\e^h = \frac{\d}{\d\tau}\e^h
\label{eq:02-55d}
\end{equation}
and integrating by parts on the right-hand side of (\ref{eq:02-55c}) finally gives
\begin{equation}
  \vec q(\tau) = \vec q^\mathrm{\,(i)}+\vec p^\mathrm{\,(i)}
  \int_0^\tau\exp\left(h(\tau')\right)\d\tau'\;.
\label{eq:02-55e}
\end{equation} 

The results for the momentum (\ref{eq:02-55b}) and the spatial trajectories (\ref{eq:02-55e}) are shown in Fig.~\ref{fig:2}. The improved trajectories (\ref{eq:02-55e}) fall between the inertial Zel'dovich trajectory and the free trajectories under the cosmological Hamiltonian (\ref{eq:02-23}).

As the inhomogeneity (\ref{eq:02-53}) shows, the gravitational potential whose gradient accelerates the particles relative to these trajectories is $g^{-1}v$ rather than $v$. According to (\ref{eq:02-16}), $g$ is growing in time; in an Einstein-de Sitter universe, $g(\tau) = (1+\tau)^{3/2}$. At late times, the potential $g^{-1}v$ is thus substantially smaller than $v$, showing that perturbations relative to the trajectories (\ref{eq:02-55e}) are much smaller at late times than those compared to the Zel'dovich trajectories.

\begin{figure*}
  \includegraphics[width=0.32\hsize]{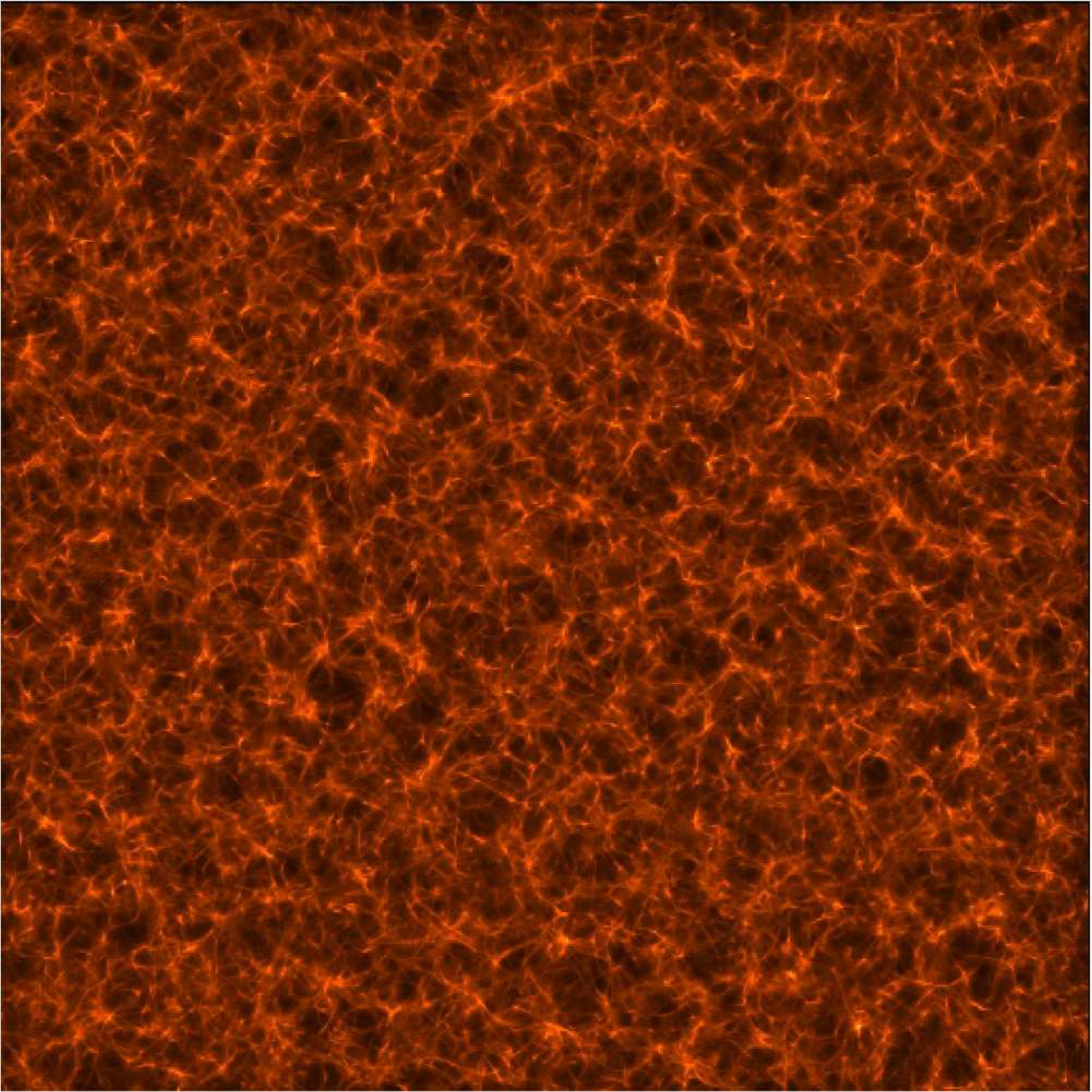}\hfill
  \includegraphics[width=0.32\hsize]{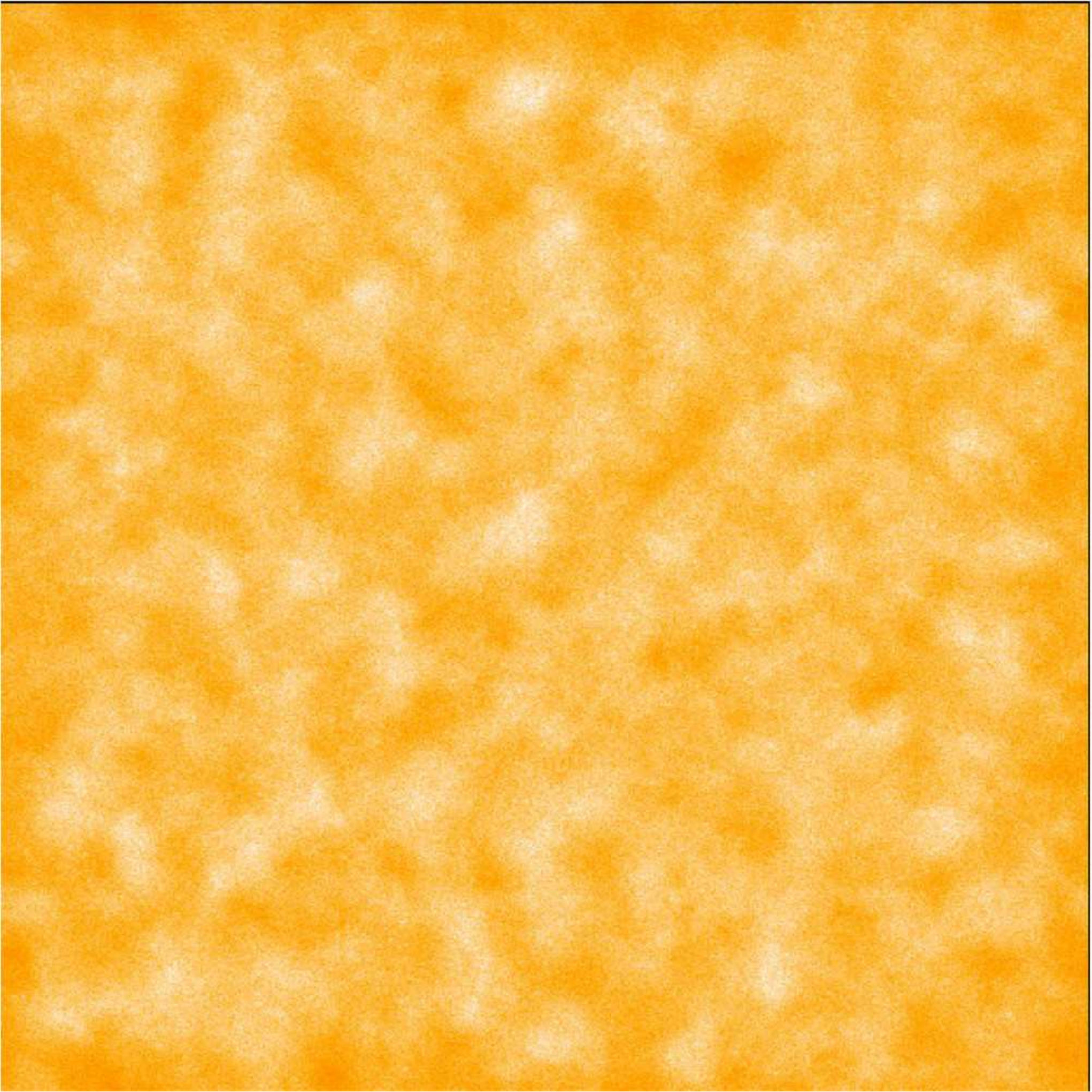}\hfill
  \includegraphics[width=0.32\hsize]{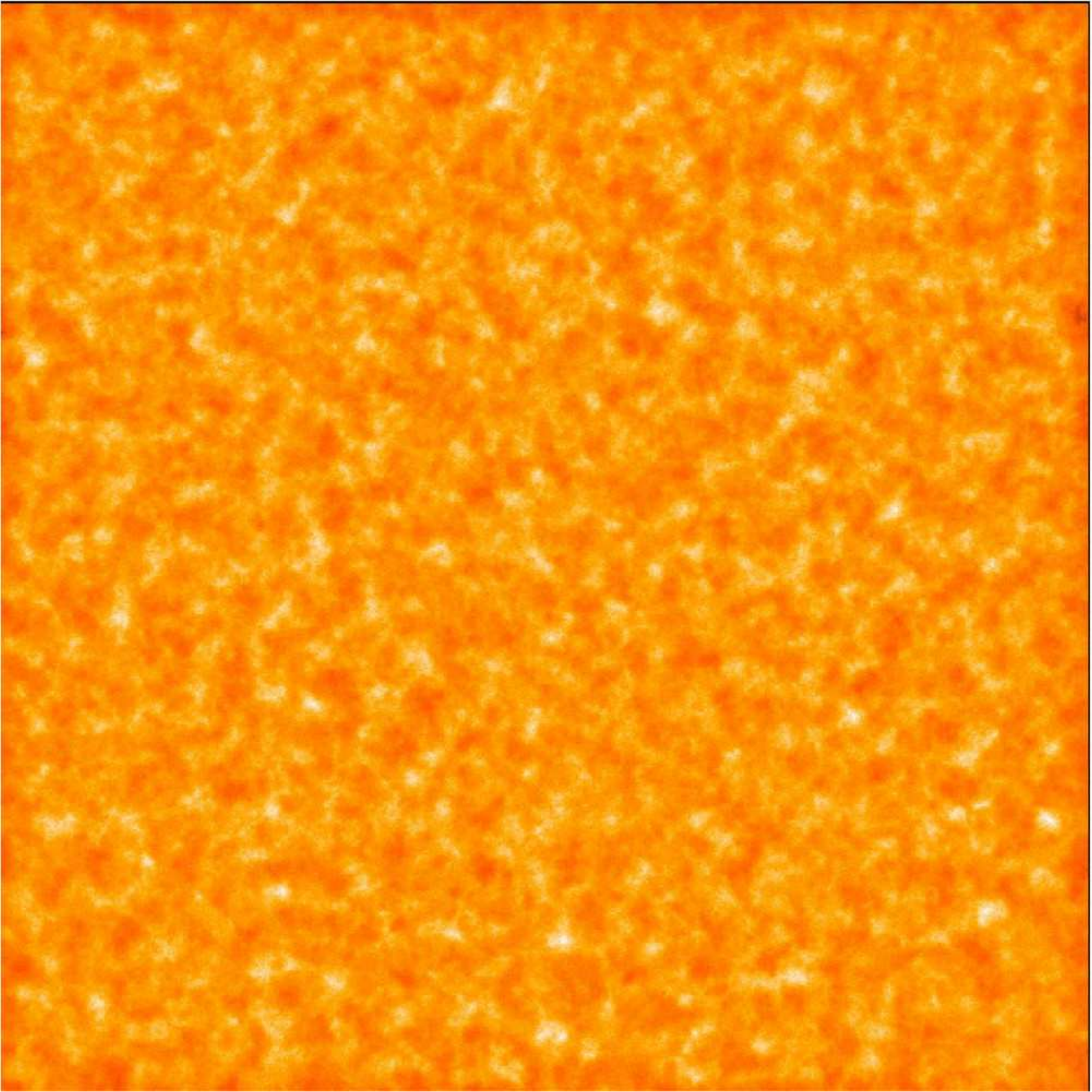}
\caption{Comparison of different approximations to large-scale structure formation. The three panels colour-encode the dark-matter density on slices through a simulation box evolved to $z = 0$ from the same CDM initial conditions in an Einstein-de Sitter universe in the adhesion approximation (left, \cite{1990ApJ...362...14N, 1990MNRAS.247..260W, 1992A&A...259..413B}), the Zel'dovich approximation (centre) and evolved with the particle trajectories (\ref{eq:02-55e}; right panel). The box size corresponds to $100\,\mathrm{Mpc\,}h^{-1}$. The improved particle trajectories result in a visibly less blurred final density configuration. The maximum density contrasts are $6.56$ in the left, $1.35$ in the central and $1.72$ in the right panel.}
\label{fig:3}
\end{figure*}

The improvement of the free trajectories given by (\ref{eq:02-55b}) and (\ref{eq:02-55e}) over the Zel'dovich approximation is illustrated in Fig.~\ref{fig:3}. Compared to the Zel'dovich trajectories, the newly derived trajectories offer several advantages. First, they lead to a substantially less blurred final density field, as Fig.~\ref{fig:3} illustrates. This was to be expected because the new trajectories improve upon the approximate inertial motion of the Zel'dovich approximation. Second, the effective gravitational potential acting relative to the new trajectories decreases with time faster by a factor of $g^{-1}$ (or $a^{-3/2}$ in Einstein-de Sitter) than the potential acting in the cosmological Hamiltonian (\ref{eq:02-23}). This implies that numerical simulations based on the new trajectories could approximate exact particle trajectories with fewer time steps. Third, while the effective gravitational potential acting on Zel'dovich trajectories, whose Fourier transform is approximated by (\ref{eq:02-52}), mixes the velocity potential $\psi$ at early times with the gravitational potential $v$ at late times, the effective potential acting on the new trajectories is simply $g^{-1}v$ and thus easier to quantify.

The faster decay with time of the effective potential $g^{-1}v$ compared to the potential $v$ is possible because part of the time dependence is moved from the potential to the free trajectories, i.e.\ to the Green's function of the free propagation.

\section{Summary}

Based on the retarded Green's functions for particles moving in a static and in an expanding space-time, we have derived two essential results:

\begin{enumerate}
  \item The effective gravitational potential experienced by particles moving on Zel'dovich trajectories acts at early cosmic times and for a short period of time only, until non-linear evolution sets in much later. At the initial time, this effective gravitational potential vanishes exactly because the initial particle velocity is constrained by the continuity equation for the density contrast. This contributes to explaining why the Zel'dovich approximation is so good even though its particle trajectories differ grossly from those expected from the Hamiltonian for point particles in an expanding space-time.
  \item The Green's function approach suggests an iteration scheme for free point-particle trajectories in an expanding space-time, which can be solved analytically. The spatial trajectories resulting from this scheme interpolate between the inertial Zel'dovich trajectories and the trajectories of free point particles with the cosmological point-particle Hamiltonian. The effective gravitational potential acting relative to these newly derived trajectories decreases with time much faster than the Newtonian potential does in an expanding space-time.
\end{enumerate}

The trajectories (\ref{eq:02-55e}) newly derived here may be useful for Lagrangian perturbation theory as well as for numerical simulations, which may be able to achieve a given spatial and temporal resolution with substantially fewer time steps because of the more rapidly decaying effective potential.

\section*{Acknowledgments}

We gratefully acknowledge inspiring and clarifying discussions with Adi Nusser, Bj\"orn M.\ Sch\"afer and Saleem Zaroubi. This work was supported in part by the Transregional Collaborative Research Centre TR~33, ``The Dark Universe'', of the German Science Foundation (DFG) and by the Munich Institute for Astro- and Particle Physics (MIAPP) of the DFG cluster of excellence ``Origin and Structure of the Universe''.

\appendix

\section{Green's functions}
\label{app:A}

\subsection{Green's function for free Hamiltonian particles}

Homogeneous equations of the type
\begin{equation}
  (\partial_t+a(t))f(t) = 0
\label{eq:02-58}
\end{equation}
are solved by
\begin{align}
  f(t) &= f_0\exp\left(-\int^t\d t'a(t')\right)\quad\mbox{or}\nonumber\\
  f(t) &= f_0\exp\left(\int_t\d t'a(t')\right)\;,
\label{eq:02-59}
\end{align}
where the first line will turn into a retarded, the second into an advanced Green's function. Since we need the retarded Green's function only, we shall drop the advanced solution right away. If $a$ is constant in time, the retarded solution simplifies to $f(t) = f_0\exp(-at)$.

By variation of constants, the retarded solution of the inhomogeneous equation
\begin{equation}
  (\partial_t+a(t))f(t) = g(t)
\label{eq:02-60}
\end{equation}
is found to be
\begin{equation}
  f(t) =  \int^t\d t'\,g(t')\exp\left(-\int_{t'}^t\d\bar t\,a(\bar t)\right)\quad(t>t')\;.
\label{eq:02-61}
\end{equation}
A retarded Green's function can be read off this result,
\begin{equation}
  g_\mathrm{R}(t,t') =  \exp\left(-\int_{t'}^t\d\bar t\,a(\bar t)\right)\theta(t-t')\;.
\label{eq:02-62}
\end{equation}
Again, this Green's function simplifies considerably if $a$ is constant in time,
\begin{equation}
  g_\mathrm{R}(t,t') =  \e^{-a(t-t')}\theta(t-t')\;.
\label{eq:02-63}
\end{equation}

The free Hamiltonian equation of motion (\ref{eq:02-1}) is of the type (\ref{eq:02-58}) with $a\to-\mathcal{K}$ and $\mathcal{K}$ constant in time. Since $\mathcal{K}^2 = 0$,
\begin{equation}
  \exp\left(\mathcal{K}t\right) = \sum_{j=0}^\infty\frac{(\mathcal{K}t)^j}{j!} = \mathcal{I}_6+\mathcal{K}t\;,
\label{eq:02-64}
\end{equation}
and the Green's functions (\ref{eq:02-63}) turns into the matrix-valued expression
\begin{equation}
  \bar G_\mathrm{R}(t,t') =
  \matrix{cc}{\mathcal{I}_3 & m^{-1}(t-t')\mathcal{I}_3 \\ 0 & \mathcal{I}_3 \\}\theta(t-t')\;.
\label{eq:02-65}
\end{equation}

For free particles in an expanding space-time, the force matrix expressing the Hamiltonian equations (\ref{eq:02-24}) simply reads
\begin{equation}
  \mathcal{K}(\tau) = \matrix{cc}{0 & g^{-1}(\tau)\mathcal{I}_3 \\ 0 & 0 \\}\;.
\label{eq:02-66}
\end{equation}
Its integral over the time $\tau'$, which we require according to (\ref{eq:02-62}), is
\begin{equation}
  \bar{\mathcal{K}}(\tau, \tau') =
  \matrix{cc}{0 & g_{qp}(\tau, \tau')\mathcal{I}_3 \\ 0 & 0 \\}
\label{eq:02-67}
\end{equation}
with
\begin{equation}
  g_{qp}(\tau, \tau') := \int_{\tau'}^\tau\frac{\d\bar\tau}{g(\bar\tau)}\;.
\label{eq:02-68}
\end{equation}
Note that $g_{qp}(\tau,\tau')$ is dimension-less because both $\tau$ and $g$ are.

Since $\bar{\mathcal{K}}^2(\tau, \tau') = 0$, we have
\begin{equation}
  \exp\left(\bar{\mathcal{K}}(\tau, \tau')\right) =
  \mathcal{I}_6+\bar{\mathcal{K}}(\tau, \tau')\;,
\label{eq:02-69}
\end{equation}
leaving us with the simple expressions
\begin{align}
  G_\mathrm{R}(\tau, \tau') &=  \matrix{cc}{\mathcal{I}_3 & g_{qp}(\tau, \tau')\mathcal{I}_3 \\ 0 & \mathcal{I}_3 \\}\theta(\tau-\tau')
  \nonumber\\
  G_\mathrm{A}(\tau, \tau') &= -\matrix{cc}{\mathcal{I}_3 & g_{qp}(\tau, \tau')\mathcal{I}_3 \\ 0 & \mathcal{I}_3 \\}\theta(\tau'-\tau)
\label{eq:02-70}
\end{align}
for the retarded and advanced Green's functions for free particles in cosmology.

In an Einstein-de Sitter model universe, $D_+ = a$, hence $f = 1$, further $\tau = a-1$ and $H = H_\mathrm{i}a^{-3/2}$, thus $g(\tau) = a^{3/2} = (1+\tau)^{3/2}$. Then, according to (\ref{eq:02-68}),
\begin{equation}
  g_{qp}(\tau, \tau') = \frac{2}{\sqrt{1+\tau'}}-\frac{2}{\sqrt{1+\tau}}\;.
\label{eq:02-71}
\end{equation}

\section{Effective Lagrange function for point particles in an expanding space-time}
\label{app:B}

This section briefly summarises the lucid treatment in \cite{1980lssu.book.....P}. For classical point particles in an expanding universe, we begin with the Lagrange function
\begin{equation}
  L(\vec r, \dot{\vec r}, t) = \frac{m}{2}\dot{\vec r}^{\,2}-m\Phi(\vec r\,)\;,
\label{eq:02-72}
\end{equation}
expressed in the physical spatial coordinates $\vec r$, with the potential $\Phi(\vec r\,)$ satisfying the Poisson equation
\begin{equation}
  \nabla_r^2\Phi = 4\pi G\rho -\Lambda
\label{eq:02-73}
\end{equation}
supplemented with the cosmological constant $\Lambda$. We introduce comoving coordinates $\vec q$ by $\vec r = a\vec q$ and transform
\begin{equation}
  L(\vec q, \dot{\vec q}, t) = \frac{m}{2}\left(\dot a^2\vec q^{\,2}+a^2\dot{\vec q}^{\,2}+2a\dot a\vec q\cdot\dot{\vec q}\,\right)-m\Phi\;,
\label{eq:02-74}
\end{equation}
where $\Phi$ is now also to be expressed in comoving coordinates.

We augment $L$ by the total time derivative of the function
\begin{equation}
  F(\vec q\,) = \frac{m}{2}a\dot a\vec q^{\,2}\;,
\label{eq:02-75}
\end{equation}
and thus obtain the effective Lagrangian
\begin{equation}
  L\to L-\frac{\d F}{\d t} = \frac{m}{2}\left(a^2\dot{\vec q}^{\,2}-a\ddot a\vec q^{\,2}\right)-m\Phi\;.
\label{eq:02-76}
\end{equation}
We further define an effective potential
\begin{equation}
  \phi = \Phi+\frac{1}{2}a\ddot a\vec q^{\,2}\;,
\label{eq:02-77}
\end{equation}
satisfying the Poisson equation
\begin{equation}
  \nabla_q^2\phi = 4\pi Ga^2\rho-a^2\Lambda+3a\ddot a
\label{eq:02-78}
\end{equation}
in comoving coordinates, and introduce the pressure-free Friedmann equation
\begin{equation}
  \frac{\ddot a}{a} = -\frac{4\pi G}{3}\bar\rho+\frac{\Lambda}{3}
\label{eq:02-79}
\end{equation}
with the mean background density $\bar\rho$ to write
\begin{equation}
  \nabla_q^2\phi = 4\pi Ga^2\left(\rho-\bar\rho\right)\;.
\label{eq:02-80}
\end{equation}
This leaves us with the effective Lagrangian
\begin{equation}
  L(\vec q, \dot{\vec q}, t) = \frac{m}{2}a^2\dot{\vec q}^{\,2}-m\phi\;.
\label{eq:02-81}
\end{equation} 

\label{lastpage}

\bibliography{../bibliography/main}

\end{document}